\documentclass[prl,twocolumn,showpacs]{revtex4}
\usepackage{graphicx,epsf}
\usepackage{bm}

\newlength{\upit}\upit=0.1truein

\newcommand{\ltappr}{{{\lower4pt\hbox{$<$} } \atop \widetilde{ \ \ \ }}}
\newlength{\bxwidth}\bxwidth=1.5 truein

\newcommand{\dg}{^{\dagger }}

\begin{document}

\title{Conserving many body approach to the fully screened, infinite $U$ Anderson model.}

\smallskip

\author{Eran Lebanon$^1$, 
Jerome Rech$^{1,2}$, P. Coleman$^1$  and Olivier Parcollet$^2$ }
\affiliation{
$^1$ Center for Materials Theory, Serin Physics Laboratory,
    Rutgers University, Piscataway, New Jersey 08854-8019, USA. \\
$^2$ Service de Physique Th\'eorique, CEA/DSM/SPhT - CNRS/SPM/URA 2306
CEA/Saclay, F-91191 Gif-sur-Yvette Cedex, FRANCE }
\begin{abstract}
Using a Luttinger Ward scheme for interacting gauge particles, we
present a conserving many body treatment of a family of fully
screened infinite $U$ Anderson models that has a smooth
cross-over into the Fermi liquid state, with a finite scattering phase
shift at zero temperature and a Wilson ratio greater than one.  We
illustrate our method, computing the temperature dependence of the
thermodynamics, resistivity and electron dephasing rate and discuss
its future application to non-equilibrium quantum dots and quantum
critical mixed valent systems.
\end{abstract}
\maketitle

The infinite $U$ Anderson model is a central element in the theory 
of magnetic moments, in their diverse manifestations within 
antiferromagnets, heavy electron systems and quantum dots. While 
the underlying physics of the Anderson impurity model is well 
understood, with a wide variety of theoretical techniques available 
for its description\cite{HewsonBook}, there are many new 
realms of physics that relate to it, such as quantum critical mixed 
valent\cite{miyake} and heavy fermion materials\cite{schof} or 
voltage biased quantum dots\cite{glazman} where our tools and 
understanding of equilibrium impurity physics are inadequate.

These considerations motivate us to seek new theoretical tools with 
the flexibility to explore the physics of the Anderson model in a 
lattice and non-equilibrium environment. One of the well-established 
tools of many body physics is the Luttinger Ward (LW) scheme\cite{LW}.  
This scheme permits a systematic construction of many-body approximations 
which preserve important interrelationships between physical variables, 
such as the Luttinger sum rule, or the Korringa relationship between spin 
relaxation and spin susceptibility,  that are implied by conservation laws.  
Recent work on the application of the LW scheme to gauge 
theories\cite{blaizot,indranil05}  provides new insight into how such 
conserving approximations can be extended to encompass 
Fermi liquids in which ultra-strong interactions are replaced by constraints 
on the Hilbert space.

In this paper we show how these insights can be applied to the infinite 
U Anderson model to obtain a description of the smooth cross-over from 
local moment to Fermi liquid behavior, in which the important conserving 
relationships between the physical properties of the ground-state, such 
as the Friedel sum rule and the Yamada-Yosida relationships between 
susceptibilities and specific heat, are preserved without approximation.  
Our method is scalable to a lattice, and can also be extended to 
non-equilibrium settings. To demonstrate the method we compute the 
temperature dependent thermodynamics, resistivity and electron dephasing 
rate in a single impurity model, and discuss how the method can be 
extended to encompass a non-equilibrium  environment. 

Our starting point is a Schwinger boson representation for the $SU (N)$
infinite $U$ Anderson model. Schwinger bosons can be used to describe 
magnetically ordered local moments, and in the limit of large $N$, 
this description becomes exact\cite{AA88}. Recent progress 
\cite{PG97,rech} has shown how this scheme can be expanded to 
encompass the Fermi liquid physics of the Kondo effect. through the 
study of a family of ``multichannel'' Kondo model, in which the number 
of channels $K$ commensurate with $n_{b}=2S$, ($K=2S$), where $S$ is 
the spin of the local moment. The corresponding infinite $U$ 
Anderson model is
\begin{eqnarray}
{\cal H} &=& \sum_{{\vec k}\nu \alpha} \epsilon_{\vec k}
c^{\dagger}_{{\vec k}\nu \alpha } c_{{\vec k}\nu \alpha}+ 
\frac{{V}}{\sqrt{N}}\sum_{{\vec k}\nu \alpha}
\left[ c^{\dagger}_{{\vec k}\nu \alpha} \chi^{\dagger}_{\nu} 
b_{\alpha} +{\rm H.c} \right] \nonumber \\
\ &+& \epsilon_0 \sum_{\sigma} b^{\dagger}_{\sigma} b_{\sigma} 
-\lambda ( Q - 2S ).
\label{model}
\end{eqnarray}
Here $c^{\dagger}_{{\vec k}\nu \alpha}$ creates a conduction
electron with momentum ${\vec k}$, channel index $\nu \in [1,K]$, 
and spin index $\alpha \in [1,N]$.  
The bilinear product $b\dg_{\alpha }\chi_{\nu }$ between a 
Schwinger boson $b^{\dagger}_{\alpha}$, and a slave fermion (or 
holon) operator $\chi_{\nu}$ creates a localized electron in the 
constrained Hilbert space, which hybridizes with the conduction 
electrons. The $\sqrt{N}$ denominator in the hybridization ensures 
a well-defined large $N$ limit. The energy of a singly occupied 
impurity $\epsilon_0$ is taken to be negative. The conserved operator 
$Q= \sum_{\alpha } b_{\alpha }^{\dagger}b_{\alpha } + \sum 
\chi^{\dagger}_{\nu } \chi_{\nu}$ replaces the no-double occupancy 
constraint of the infinite $U$ Anderson model, where $Q=2S=K$ is 
required for perfect screening.

The characteristic ``Kondo scale'' of our model is 
\begin{equation}
T_K = D \left( \frac{\Gamma}{\pi D} \right)^{K/N} 
{\rm exp} \left\{ -\frac{\pi|\epsilon_0|}{\Gamma} \right\},
\label{Tk}
\end{equation}
where $D$ is the conduction electron half bandwidth, $\Gamma = \pi 
\rho {{V}}^2$ is the hybridization width and $\rho $ is the
conduction electron density of states. This relationship follows 
from general leading logarithmic scaling of the model. 
The non-trivial prefactor in the Kondo temperature depends on the 
third order  terms in the beta function. 

\begin{figure}[tb]
\centerline{ \includegraphics[width=80mm]{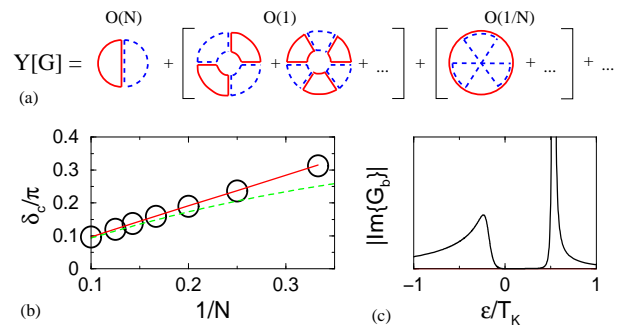} \vspace{-5pt} 
}
\caption{
        (a)LW functional grouped in powers of $1/N$. Solid line - 
        Schwinger bosons, dashed lines - holons, 
	double lines - conduction electrons. Loops contribute 
	$O(N)$, vertices $1/\sqrt{N}$.
	(b) Circles: phase shifts extracted from the calculated 
	t-matrices, for $K=1$, $n_{\chi}\approx 0.04$.
	Solid line stands for $(K-n_{\chi})/(NK)$, dashed line 
	shows the phase shift without imposing self 
	consistency on the conduction electron propagators. 
	(c) The boson spectrum is gapped at low temperatures.
	}
\label{fig:fig1}
\end{figure}

The next step is to construct the LW functional $Y[{ G}]$. The 
variation of $Y[G]$ with respect to the full Green's functions 
${G}_{\zeta}$ of the conduction, Schwinger boson and slave fermion
fields, $\Sigma_{\zeta} = \delta Y / \delta { G}_{\zeta}$, 
self-consistently determines the self-energies $\Sigma_{\zeta}$ of 
these fields\cite{blaizot,indranil05}. $Y[G]$ is equal to the sum 
of all two-particle irreducible free energy Feynman diagrams, 
grouped in powers of $1/N$ (Fig.\ref{fig:fig1}(a)).
In 
our procedure we neglect all but the leading order $O (N)$ term in the LW
functional, and then impose the full self-consistency generated by
this functional. 

In the  formal large $N$ limit, the conduction electron 
self-energy is of order $O (1/N)$, and at first sight, should be 
neglected. 
However, the effects of these terms on the free energy 
are enhanced by the number of channels $K$ and spin 
components $N$, to give a leading order $O (N)$ contribution to the 
free energy: it is precisely these terms that produce the Fermi 
liquid behavior\cite{rech}. The complex issue of exactly when the 
conduction electron self-energies should be included is naturally 
resolved by  using the leading order LW functional, subsequently 
treating $1/N$ as a finite parameter, using the conduction electron 
self-energies inside the self-consistency relations. The explicit 
expressions for the self-energies are then
\begin{eqnarray}  
\Sigma_{\chi}(\tau) &=& \ \ \ \ \ \ \ \ \ \ \ 
{{V}}^2 G_b(\tau) G_c(-\tau) \nonumber \\ 
\Sigma_b (\tau) &=& -(K/N) {{V}}^2 G_{\chi}(\tau) G_c(\tau) \nonumber \\ 
\Sigma_c (\tau) &=&  \ \  (1/N)\ {{V}}^2 G_b(\tau) G_{\chi}(-\tau)
\label{mPG}
\end{eqnarray}
where $G_b$, $G_{\chi}$ and $G_c$ are the fully dressed imaginary time
Green's functions of the boson, holon and conduction electrons
respectively.  Equations (\ref{mPG}) are solved self consistently,
adjusting the chemical potential $\lambda$ to satisfy the averaged
constraint $\langle Q \rangle =K$.  

A key step in the derivation of the Ward identities 
is the presence of a scale in the excitation spectrum, which manifests itself
mathematically in our ability to replace
Matsubara summations by integrals at low temperatures:
\[
T\sum_{i\omega_{n}} \rightarrow \int \frac{d\omega}{2\pi}
\]
This transformation is the key to derivations of the Friedel sum
rule, the Yamada-Yosida and Shiba 
relationships between susceptibilities, specific heat and spin
relaxation rates\cite{langreth,yamada,shiba}. 
There is  long history of attempts to develop conserving
approximations to the infinite U Anderson model
\cite{kuromoto,coleman}, but each has been thwarted 
by the development of scale-invariant 
X-ray singularities in the gauge or slave particle spectra for which
the above replacement is illegal.  In the multi-channel approach,
the gauge particles - the formation of a stable Kondo singlet when $2S=K$
leads to the formation of a gap in the spinon and holon 
spectrum [see Fig. \ref{fig:fig1}(c)], 
which permits a Fermi liquid spectrum to 
develop at lower energies. This feature permits {\sl all} of the 
manipulations previously carried out on the finite U Anderson model 
to now be extended to infinite $U$.  Moreover, these relationships 
are all satisfied by the use of the leading LW functional.

There are many important conserving relationships that are 
preserved, provided the spinons and holons develop a gap. 
The first of these is the Friedel sum rule (or in the lattice, 
the Luttinger sum rule), according to which the sum of the 
conduction electron phase shifts
must be equal to the total charge $ K-n_{\chi }$ on the impurity, 
whith $n_{\chi}$ as the ground-state holon occupancy, so
\begin{equation}
\delta_c = \frac{\pi}{N} \frac {K -  n_{\chi} }{K} + O\left( 
\frac{T_K}{D}\right) ,
\label{phase_shift}
\end{equation}
Fig. 1. contrasts the dependence of phase shift obtained from the
t-matrix in our  conserving scheme, with that obtained using 
$\Sigma_c$ from the leading order $1/N$ expansion. 
The relationship (\ref{phase_shift}) is obeyed at each 
value of $N$ in our approximation. 

The quasiparticle spectrum in the Fermi liquid 
ground-state is intimately related to the conduction electron phase
shifts via  Nozi\'eres Fermi liquid theory\cite{nozieres}. The change 
in the scattering phase shift in response to an applied field, or a 
change in chemical potential, is directly related to the  spin, 
charge and channel susceptibilities, which in turn,  leads to 
a generalized  ``Yamada-Yosida identity'', distributing the $NK$
degrees of freedom among the spin, flavor and charge sectors: 
\begin{equation}
NK\frac{\gamma}{\gamma^0} = 
K\frac{N^2 -1}{N+K} \frac{\chi_s}{\chi_s^0} + 
N\frac{K^2-1}{K+N} \frac{\chi_f}{\chi_f^0} 
+\frac{\chi_c}{\chi_c^0}.
\label{WilsonRatio}
\end{equation}
Here $\gamma$ stands for the specific heat coefficient $\gamma=C/T$, 
and $\chi_{s}$, $\chi_f$ and $\chi_c$ stand for the spin, flavor and 
charge susceptibilities. $\chi_{s,f,c}^{0}$ denotes the corresponding
susceptibilities in the absence of the impurity.
Our  identity (\ref{WilsonRatio}) 
reduces to the known results for the Wilson ratio in the one channel
model\cite{yamada} and the Kondo 
limit\cite{JAZ98}. 
Since our scheme preserves the
Friedel sum rule for each channel, it automatically satisfies
this relationship. 

\begin{figure}
\centerline{
\includegraphics[width=80mm]{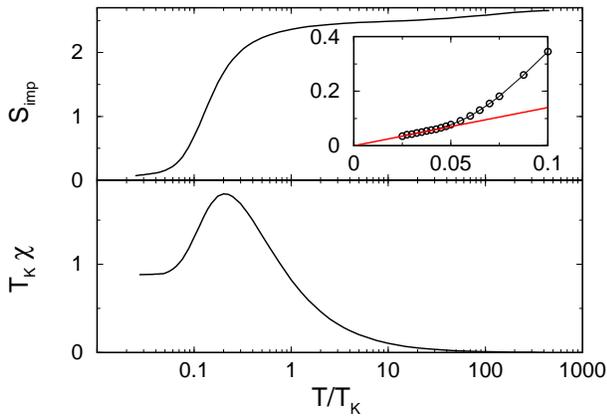}
}\vspace{-5pt} \caption{
        Upper pannel: The impurity contribution 
        to the entropy vs. temperature. Inset: at low temperatures 
	the entropy becomes linear and $\gamma$ saturates. Lower 
	pannel: impurity susceptibility vs. temperature.  
        For parameters see Fig.\ref{fig:fig2}.
        }
\label{fig:fig3}
\end{figure}

Various thermodynamic quantities can be extracted from the
numerical solution of the self-consistency equations.
The development of a fully quenched  Fermi liquid state is
succinctly demonstrated by calculating the entropy, using a 
formula of Coleman, Paul and Rech\cite{indranil05}
\begin{equation}
S=-
{\rm Tr}_{\zeta}
 \int d\epsilon 
\frac{dn_{\zeta}}{dT} \left[ {\rm Imln}\bigl(-G_{\zeta}^{-1}\bigr)+
G'_{\zeta} \Sigma''_{\zeta} \right],
\end{equation}
where the trace denotes a sum over the spin and channel indices 
of the various fields and $n_{\zeta}$ is the distribution function 
(Fermi for $\zeta=c,\chi$ and Bose for $\zeta=b$). $G_{\zeta}$ and 
$\Sigma_{\zeta} $ respectively denote the retarded Green functions 
and self energies. The single and double primes mark the real and 
imaginary part respectively. The impurity contribution to the 
entropy $S_{\rm imp} = S - S_0$, where $S_0$ is the entropy of the 
bare conduction band, is shown in the upper pannel of Fig. \ref{fig:fig3}.
An inspection of low temperatures shows a 
linear behavior of the impurity contribution to the entropy. The 
latter marks the saturation of the specific heat 
coefficient $\gamma_{\rm imp}=\partial S_{\rm imp}/\partial T$ at 
low temperatures and the formation of a local Fermi liquid.

To calculate the spin susceptibility, we couple a small magnetic field
to the magnetization, defined as 
\[
M= \sum_{\alpha }{\hbox{sgn}} (\alpha-
\frac{N+1}{2})\biggl(
b\dg _{\alpha}b_{\alpha}
+\sum_{\vec k,\nu
} c\dg _{\vec k \nu \alpha }c_{\vec k \nu
\alpha }  \biggr).
\]
The impurity magnetization is obtained by subtracting the magnetization
of the free conduction sea from $M$. 
When we apply a field, we must keep track of the vertex
corrections that arise from the field-dependence of the
self-energies. These vertex functions are essential to maintain the
conservation laws. We compute the magnetic field vertex functions by 
iterating the solutions to self-consistency in a small magnetic 
field\cite{LSZ01}.
Figure (\ref{fig:fig3}) shows the impurity susceptibility $\chi(T)$. 
The susceptibility is peaked slightly below the Kondo temperature and 
saturates to a constant at a lower temperature. 

One of the important conservation laws associated with the
conservation of spin, is the Korringa Shiba relation between the
dynamical and static spin susceptibility. On the assumption that the
gauge particles are gapped, the LW derivation carried
out by Shiba on the finite $U$ Anderson model, some thirty years ago,
can be simply generalized to the infinite $U$ model, 
to obtain 
$\left. \chi''(\omega)/\omega\right|
_{\omega = 0}=  (N\pi/2K) (\chi/N)^2$. 
This relationship guarantees that the power-spectrum of the
magnetization is linear at low frequencies, ensuring that the
spin response function decays as $1/t^{2}$ in time. Remarkably,  
a Fermi liquid, with slow gapless spin excitations is sandwiched beneath
the spinon-holon continuum. 

\begin{figure}
\centerline{
\includegraphics[width=80mm]{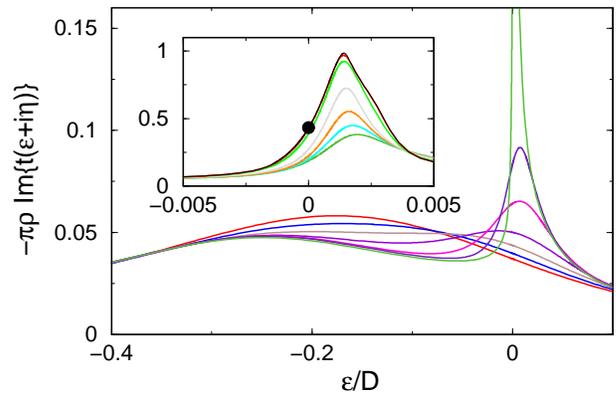}
}\vspace{-5pt} \caption{
        t-matrix for $\epsilon_0/D=-0.2783$, 
        $\Gamma/D=0.16$, $N=4$ and $K=1$ ($T_K/D=0.002$). 
	Main frame: T/D=0.1, 0.08, 0.06, 0.04, 0.02, and 0.01. 
	Inset: T/$10^{-4}$D=10, 8, 6, 4, 2, 1, and 0.5.
	The dot marks ${\rm sin}^2 \{ \pi (K-n_\chi)/NK \} $
        }
\label{fig:fig2}
\end{figure}

We now turn to a discussion of the electron scattering off the
impurity, which is determined by the conduction electron t-matrix, 
$t (z)= \frac{\Sigma_{c}}{1 - G^{(0)}_{c} \Sigma_{c}}$, where $G^0_{c}
(z)= \sum_{\vec{ k}} (
z-\epsilon_{\vec{k}})^{-1}$ 
is the bare local conduction electron propagator. Various important
physical properties, such as the impurity resistivity and the electron
dephasing rate in the dilute limit can be related to this quantity. 
Figure \ref{fig:fig2} shows the imaginary part of 
$t (\epsilon+i\eta)$: as temperature is lowered towards the Kondo 
temperature, a resonance peak starts to develop around the Fermi 
energy. The peak continues to evolve with temperature and finally 
saturates a decade below the Kondo temperature. The fully developed 
resonance is pinned to the Langreth sum-rule\cite{langreth} value of 
${\rm sin}^2 \delta_c/\pi\rho $, at the Fermi energy presented by the 
dot in the inset of Fig.\ref{fig:fig2}. 

\begin{figure}
\centerline{
\includegraphics[width=80mm]{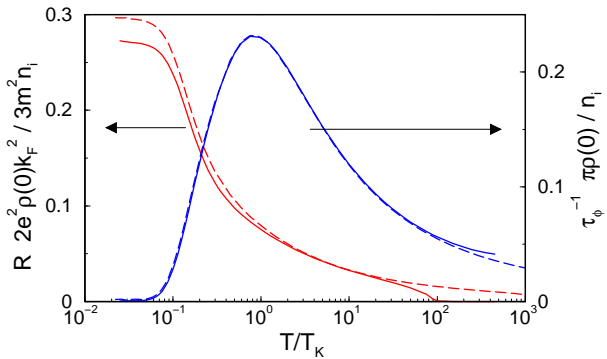}
}\vspace{-5pt} \caption{
	The impurity contribution to the resistivity $R_i$ 
        and dephasing rate $\tau_{\phi}^{-1}$ as a function of 
	temperature. Solid lines: the parameters used in
	Fig. \ref{fig:fig2}, dashed lines: $\Gamma=0.1 D$.}
\label{fig:fig4}
\end{figure}

The impurity contribution to the resistivity $R_i$ is related to a 
thermal average of the $t-$ matrix, 
\begin{equation}
R_i =  \frac{3m^2n_i}{2e^2\rho k_F^2} \left\{
\int d\epsilon \left( -\frac{\partial f}{\partial \epsilon} 
               \right) \left| {\rm Im} t(\epsilon+i\eta) 
	               \right|^{-1} \right\}^{-1}.
\end{equation}
where 
$k_F$ is the Fermi wavelength and $f$ is the Fermi distribution function. 
Figure \ref{fig:fig4} shows $R_i(T)$. The resistivity increases 
as the temperature decreases, 
saturating at a value determined by the scattering phase shift. 

As a final application of the t-matrix, we compute the electron dephasing 
rate $\tau_{\phi}^{-1}$ that controls the field dependence of weak electron 
localization. In the cross-over to the Fermi liquid state, the formation of 
the Kondo resonance gives rise to a peak in the inelastic 
scattering\cite{Zarand04} and the 
electron dephasing rate \cite{dephasing}. In the dilute limit, the impurity 
contribution to the dephasing rate is given by 
$\tau_{\phi}^{-1} = 2 n_i [ -{\rm Im}t - \pi \rho |t|^2 ]$  
\cite{MACR05,Zarand04,Bauerle05}.  
Fig. \ref{fig:fig4} shows the dephasing rate on the Fermi surface
computed from $t$, showing how the scattering scales with the Kondo 
temperature $T_{K}$ in the Kondo regime of the model.

Our calculation of the dephasing rate shows that our method captures
the cross-over into the coherent Fermi liquid. However, it 
also highlights a shortcoming that we hope to address in
future work. At low temperatures, the electron dephasing rate has a
quadratic dependence on temperature. 
The electron-electron scattering diagrams responsible for these
processes involve an internal loop of gauge particles, which only
enters in the subleading $O (1)$ terms of the LW functional shown in
Fig. 1(a), which are absent from the current work. 
On the other hand, the low temperature, on-shell value of the
conduction electron vertex is directly related to $\gamma$ by 
Ward identities \cite{MACR05}, providing a possible future simplification
for treating these processes.

There are many directions for future development. We are particularly 
interested in the extension to non-equilibrium quantum dots\cite{glazman}. 
The theoretical understanding of the Kondo physics of quantum dots at finite 
voltage bias is still evolving. An extension of our method to a voltage 
biased infinite Anderson model can be made using a Keldysh generalization 
of the self-consistency equations\cite{anna}. Many features of our 
approach, its conserving properties, its inclusion of scaling properties 
up to third order in the beta function,  and its systematic 
incorporation of spin relaxation effects on the boson lines, suggest
that this will be a robust method to examine how spin and electron
dephasing effects evolve with voltage in quantum
dots. These methods can also be used to study the voltage dependence 
\cite{eranpiers} of the distribution function and 
noise in magnetically doped mesoscopic wires\cite{relaxation}.

One of the striking  features of the Schwinger boson approach to the
Anderson and Kondo models, is the co-existence of a gapless Fermi liquid,
sandwiched at low energies between a gapped spinon and holon fluid.
The gap in the spin-charge decoupled excitations appears intimately 
linked to the development of the Fermi liqid Ward identities. 
The methods presented in this paper can be scaled 
up to describe the dense Anderson and Kondo lattice models, where the 
Friedel sum rule is replaced by the Luttinger sum rule\cite{indranil05}. 
In the lattice, the $U (1)$ local symmetry of the impurity model will in 
general be replaced by a global $U (1)$ symmetry  associated with the 
pair-condensed Schwinger bosons\cite{senthil}. In this setting, the 
gapped spinons and holons are propagating excitations, whose gap is 
fundamental to the large Fermi surface. It is this very gap that we expect 
to collapse at a quantum critical point. The current work, which includes 
the effects of valence fluctuations, provides a powerful framework for 
examining this new physics. 

This work was supported by DOE grant number
DE-FE02-00ER45790 and an ACI grant of the French Ministry of Research.
We are grateful to Natan Andrei and 
Gergely Zarand for discussions related to this work.


\end{document}